# Throughput Computation in CSMA Wireless Networks with Collision Effects

Cai Hong Kai, Soung Chang Liew
Department of Information Engineering, The Chinese University of Hong Kong
Email: caihong.kai@gmail.com, soung@ie.cuhk.edu.hk

Abstract

It is known that link throughputs of CSMA wireless networks can be computed from a time-reversible Markov chain arising from an *ideal CSMA network* model (ICN). In particular, this model yields general closed-form equations of link throughputs. However, an idealized and important assumption made in ICN is that the backoff countdown process is in "contiuous-time" and carrier sensing is instantaneous. As a result, there is no collision in ICN. In practical CSMA protocols such as IEEE 802.11, the stations count down in "mini-timeslot" and the process is therefore a "discrete-time" process. In particular, two stations may end their backoff process in the same mini-timeslot and then transmit simultaneously, resulting in a packet collision. This paper is an attempt to study how to compute link throughputs after taking such backoff collision effects into account. We propose a *generalized ideal CSMA network* model (GICN) to characterize the collision states as well as the interactions and dependency among links in the network. We show that link throughputs and collision probability can be computed from GICN. Simulation results validate GICN's accuracy. Interestingly, we also find that the original ICN model yields fairly accurate results despite the fact that collisions are not modeled.

## I Introduction

With the widespread deployment of IEEE 802.11 networks, it is common today to find multiple wireless LANs co-located in the neighborhood of each other. The multiple wireless LANs form an overall large network whose links interact and compete for airtime using the carrier-sense multiple access (CSMA) protocol. When a station hears its neighbors transmit, it will refrain from transmitting in order to avoid packet collisions.

For analytical purposes, the carrier sensing relationships among the links are typically captured using a contention graph. The links are modeled by vertices of the graph, and an edge joins two vertices if the transmitters of the two associated links can sense each other. Since different links may sense different subsets of other links, the links may experience different throughputs.

Ref. [1] presented an analytical model, Ideal CSMA Network (ICN), to study the behavior of CSMA networks given their contention graphs. It was shown that the throughputs of links can be computed from the stationary probability distribution of the states of a continuous-time Markov chain, even though the process is not memoryless. The same model has been used in several prior works [2]

[3], assuming that the backoff and transmission time are exponentially distributed. An important contribution of [1] is the removal of this assumption, making the ICN model applicable to a practical CSMA wireless network. Recently, an elegant distributed adaptive CSMA algorithm is proposed based on the ICN model to achieve the optimal throughput in CSMA wireless networks [4] [5].

Although the ICN model has captured the main features of the CSMA protocol, an important simplifying assumption made in order to maintain analytical tractability is that there is no collision in the network. In particular, it is assumed that carrier sensing is perfect and instantaneous. The goal of ICN is to capture the "first-order" interactions among the links due to their carrier-sensing relationships as modeled by the contention graph. Collisions are treated as a "second-order" effect and modeled away in ICN. An outstanding issue is how to make ICN more accurate by incorporating the effects of collisions.

In practical CSMA protocol such as IEEE 802.11, time is divided into discrete minislots, and collisions happen if multiple conflicting links count down to zero in the same timeslot in their backoff process and then transmit simultaneously[1]. When a collision occurs, all the involved links lose their packets and will try to retransmit later.

In this paper, we first propose a generalized ideal CSMA network model (GICN) to analyze the effects of collisions due to simultaneous transmissions. GICN can be viewed as a perturbation-analytical model of ICN. Based on the GICN model, the link throughputs and collision probability can be computed. Simulation results show that the GICN model has high accuracy. Interestingly, we find that the effect of collisions is not significant as far as the link throughputs are concerned. That is, the original ICN model yields good approximations even though it does not capture the collision effect.

## II ICN Model and Its Equilibrium Analysis

To build up the background for the later sections, we briefly review the ICN model and its equilibrium analysis here.

*A. ICN Model*

In ICN, the carrier sensing relationship among links is described by a contention graph $G = (V, E)$. Each link is modeled as a vertex $i \in V$. Edges, on the other hand, model the carrier-sensing relationships among links. There is an edge $e \in E$ between two vertices if the transmitters of the two associated links can sense each other. In this paper we will use the terms "links" and "vertices" interchangeably.

At any time, a link is in one of two possible states, active or idle. A link is active if there is a data transmission between its two end nodes. Thanks to carrier sensing, any two links that can hear each other will refrain from being active at the same time. A link sees the channel as idle if and only if none of its neighbors is active.

---

[1] The "hidden-node" phenomenon can also cause packet collisions. That is, the transmitters of two links cannot hear the activities of each other, but a collision occurs if the two transmissions overlap. The throughput analysis with collisions incurred by "hidden-node" is left as future work in this paper. Alternatively, we can design protocols, such as the algorithms proposed in [6], to remove the hidden-node phenomenon.

In ICN, each link maintains a *backoff timer*, $C$, the initial value of which is a random variable with an *arbitrary* distribution $f(t_{cd})$. The timer value of the link decreases in a continuous manner with $dC/dt = -1$ as long as the link senses the channel as idle. If the channel is sensed busy (due to a neighbor transmitting), the countdown process is frozen and $dC/dt = 0$. When the channel becomes idle again, the countdown continues and $dC/dt = -1$ with $C$ initialized to the previous frozen value. When $C$ reaches 0, the link transmits a packet. The transmission duration is a random variable with *arbitrary* distribution $g(t_{tr})$. After the transmission, the link resets $C$ to a new random value according to the distribution $f(t_{cd})$, and the process repeats. We define *the access intensity* of a link as the ratio of its mean transmission duration to its mean backoff time: $\rho = E[t_{tr}]/E[t_{cd}]$.

Let $s_i \in \{0,1\}$ denote the state of link $i$, where $s_i = 1$ if link $i$ is active (transmitting) and $s_i = 0$ if link $i$ is idle (actively counting down or frozen). The overall **system state** of ICN is $s = s_1 s_2 ... s_N$, where $N$ is the number of links in the network. Note that $s_i$ and $s_j$ cannot both be 1 at the same time if links $i$ and $j$ are neighbors because (i) they can sense each other; and (ii) the probability of them counting down to zero and transmitting together is 0 under ICN (because the backoff time is a continuous random variable).

The collection of feasible states corresponds to the collection of independent sets of the contention graph. An independent set (IS) of a graph is a subset of vertices such that no edge joins any two of them [7].

As an example, Fig. 1(a) shows the contention graph of a network consisting of four links. In this network, link 1 only senses link 2 while links 2, 3 and 4 can hear each other. Fig. 1(b) shows the associated state-transition diagram under the ICN model. To avoid clutters, we have merged the two directional transitions between two states into one line in Fig. 1(b). Each transition from left to right corresponds to the beginning of the transmission of one particular link, while the reverse transition corresponds to the ending of the transmission of that link. For example, the transition $1000 \rightarrow 1010$ is due to link 3's beginning to transmit; the reverse transition $1010 \rightarrow 1000$ is due to link 3's completing its transmission.

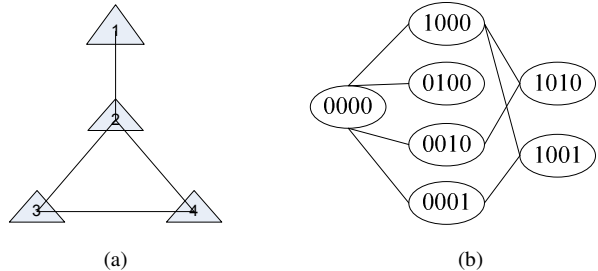

Fig. 1. (a) An example contention graph and (b) its state-transition diagram.

*B. Equilibrium Analysis*

This part is a quick review of the result in [1], and the reader is referred to [1] for details. If we assume that the backoff time and transmission time are exponentially distributed, then $s(t)$ is a

*time-reversible* Markov process. For any pair of neighbor states in the continuous-time Markov chain, the transition from the left state to the right state occurs at rate $\lambda = 1/E[t_{cd}]$, and the transition from the right state to the left state occurs at rate $\mu = 1/E[t_{tr}]$.

Let $\mathcal{S}$ denote the set of all feasible states, and $n_s$ be the number of transmitting links when the system is in state $s = s_1 s_2 ... s_N$. The stationary distribution of state $s$ can be shown to be:

$$P_s = \frac{\rho^{n_s}}{Z} \quad \forall s \in \mathcal{S}, \quad \text{where} \quad Z = \sum_{s \in \mathcal{S}} \rho^{n_s} \tag{1}$$

The fraction of time during which link $i$ transmits is $th_i = \sum_{s:s_i=1} P_s$, which corresponds to the normalized throughput of link $i$.

Ref. [1] showed that (1) is in fact quite general and does not require the system state $s(t)$ to be a Markov process. In particular, (1) is insensitive to the distribution of the transmission duration $g(t_{tr})$, and the distribution of the backoff duration $f(t_{cd})$, given the ratio of their mean $\rho = E[t_{tr}]/E[t_{cd}]$.

Note that $Z$ in (1) is a weighted sum of independent sets of $G$. In statistical physics, $Z$ is referred to as the partition function and the computation of $Z$ is the crux of many problems, which is known to be NP-hard [7].

For the case where different links have difference access intensities, (1) can be generalized by replacing $\rho^{n_s}$ with the $\prod_{i:s_i=1 \text{ in } s} \rho_i$, where $\rho_i$ is the access intensity of link *i*.

## III GICN Model and Assumptions

This section describes our generalized ideal CSMA model (GICN) and the assumptions made in the analysis of Section IV.

Different from ICN where both the backoff time and transmission time are *continuous* variables, in GICN time is divided into mini-timeslots. The backoff countdown timer is an *integer* variable which is uniformly distributed between [0, CW], where CW is the contention window. The timer value of the link decreases by one for each timeslot the link senses the channel as idle. Furthermore, we assume that the transmission duration is a fixed value equal to $E[t_{tr}]$ (As can be seen later, this assumption is important in our analysis.). That is, the access intensity of a link is $\rho = 2E[t_{tr}]/CW$.

The feasible state of ICN is the independent set of the contention graph. In GICN, multiple links can countdown to zero in the same timeslot and then transmit together. That is, the states of two neighbor links, $s_i$ and $s_j$ can both be 1 at the same time. In this case, we say that a collision happens and the two packets collide (in this paper we do not consider the "signal capture" effect [8]).

We summarize the assumptions made in GICN as follows:
1) The contention window, $CW$, is not doubled upon collisions.

2) The transmission time is of fixed length of $E[t_{tr}]$.
3) The remaining countdown time of each link is independent of each other when collision events are taken into account.
4) Only collisions in which two neighbor links transmit together are considered. That is, we do not consider the collisions in which more than two neighbor links transmit at the same time.

Ref. [2] made the same assumption as 1) and studied the throughput computation and fairness issues in CSMA wireless networks. Note that [2] did not consider collision events. This paper studies the throughput computation of links in CSMA wireless networks with collisions. To justify this assumption, we will show the impact of contention window doubling is not significant through simulation results in Section V.

Note from 2) that in GICN, the links involved in a collision are "synchronized". That is, the links involved in a collision must begin transmitting in the same timeslot and will end transmitting in the same timeslot after $E[t_{tr}]$ timeslots. It can be shown that, under this assumption, the system process is still time-reversible.

Ref. [9] made the same assumption as 1), 2) and 3) to design CSMA-based scheduling algorithms. Different from [9] in which new distributed scheduling algorithm is designed, this paper studies the existing CSMA protocol such as IEEE 802.11 networks.

## IV Throughput Distribution Computation under GICN

We conduct perturbation on ICN's computation of link throughputs. The basic idea is as follows: in the state-transition diagram of ICN, we can figure out the states starting from which a collision may occur. For each such state, we could carefully identify the transition rates between these collision-related states. Then we can compute the stationary probability of each state, obtain the normalized link throughputs and the collision probability of each link.

In Part A below, we calculate the conditional collision probability beforehand, which will be used in the analysis of Part B.

*A. Conditional Collision Probability*

For each collision caused by simultaneous transmissions, the multiple neighbor links involved in the collision must be actively counting down before the collision occurs.

Consider the collision probability of a particular link conditioned on that the link has $n$ actively-countdown neighbors. In our model the backoff time of each link is uniformly distributed between [0, CW]. If we treat transmission slot as one countdown backoff slot, then the average distance between two transmission time slots is $\frac{CW}{2}$, and the transmission probability of an actively counting down station over each timeslot is $\frac{1}{CW/2+1} = \frac{2}{CW+2}$.

Take the typical value of contention window in IEEE 802.11b network, $CW = 31$, as an example. The average distance between two transmission time slots is 15.5, and the transmission probability of an actively counting down station is over each timeslot 1/16.5. The probability of collision is then $1-(15.5/16.5)^n$.

Conditioned on that an actively-countdown link has $n$ actively-countdown neighbors, the collision probability of this link, $q_n$, is

$$q_n = 1 - \left(1 - \frac{2}{CW+2}\right)^n = 1 - \left(\frac{CW}{CW+2}\right)^n \qquad (2)$$

**Remark:** From (2) we have $q_1 = \frac{2}{CW+2}$. Filling in the typical value of $CW = 31$, $q_1 = 0.0607$. When $q_1$ is small, we can approximate (2) as

$$q_n = 1 - (1-q_1)^n \approx nq_1 \qquad (3)$$

We note that the approximation made in (3) is important in the construction of state-transition diagram in Part B.

## B. Perturbation on ICN's Computation

Next we illustrate our perturbation on ICN's computation. We use the two-link network as the illustrating example. The state transition is shown in Fig. 2. State "$**_{00}$" is the collision state (i.e., it is the state 11 in this example). Since the transmission time is fixed, starting from state "$**_{00}$", the system will evolve back to state "00" directly.

**Procedure of Conducting Perturbation on ICN's Computation**

**Step 1:** From the state-transition of ICN, we identify the states in which collisions may occur. Denote this subset of states by $M$. For each $s \in M$, the collision states connected to $s$ are represented by a single state, denoted by $**_s$. State $**_s$ includes all possible states in which two-link collision occurs. The probability of a collision state $**_s$ is the sum of the probabilities of these states.

In Fig.2, states 01 and 10 cannot lead to a "collision" state. Starting from 00, there could be a state 11, in which the two links collide. That is, $M = \{00\}$ and $**_{00} = \{11\}$.

**Step 2:** For each state $s$ in $M$, identify the transition rates to/from its neighbor states. Note that the transition rate from "1" to "0" is fixed to $\mu$ no matter whether the transmission collides with a neighbor's packet. Also, the transition rate from state $**_s$ to $s$ is $\mu$.

To identify the transition rates of the collision-related states, we need to compute the conditional collision probability first. For each state $s$ in $M$, we identify the collision probability of each link conditioned on that the current state is $s$, using the results in Part A.

In our example of Fig. 2, given state $s = 00$, the collision probability of both links 1 and 2 is $q_1$. The transition rate from 00 to ** is then $q_1\lambda$; the transition rate from 00 to 01 and 10 becomes

$(1-q_1)\lambda$. We note that the transition rate leaving state 00 is a bit smaller than $2\lambda$ due to collision effects.

In general, we can count the number of neighbor links of link $i$ which are actively counting down in state $s$. Invoking (2) or (3), we can obtain the collision probability of link $i$. The transition rate from state $s$ to the collision state $**_s$ is equal to $q_n$, where $n$ is the number of edges joining the actively-countdown links in state $s$.

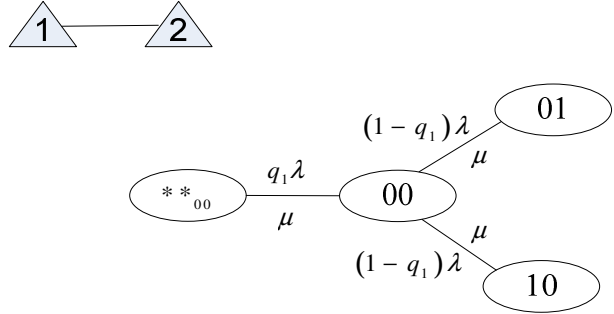

Fig. 2 Two-link network and the associated state-transition diagram with collisions

**Step 3:** Assuming that the stationary probability of GICN has the product form as in ICN (Note that local balance still holds in the generalized state-transition diagram because it can be shown that the system process is time-reversible), we can compute the stationary probability distribution of states in GICN. Based on this stationary distribution, we can compute: i) the normalized throughputs of links $Th_i = \sum_{s:s_i=1} P_s$ and ii) the collision probability of each link: $p_i = \dfrac{\Pr\{\text{collision states in which link } i \text{ is collided}\}}{\Pr\{\text{all states in which link } i \text{ is active}\}}$.

In our example of Fig. 2, we have $Th_1 = Th_2 = \dfrac{(1-q_1)\rho}{1+2\rho-q_1\rho}$ and $p_1 = p_2 = q_1$. Given a typical value of $\rho = 83/15.5$, the normalized throughput of both links is 0.4418, collision probability of both links is $q_1 = 0.0607$. MATLAB simulation validated this result.

We use two more examples to demonstrate the procedure:

1) A three-link network

For three link case, the state transition is shown in Fig.3. "$***_{000}$" represents all possible collision states. In Fig. 2, $q_1, q_2$ is calculated from Eq.2.

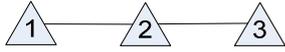
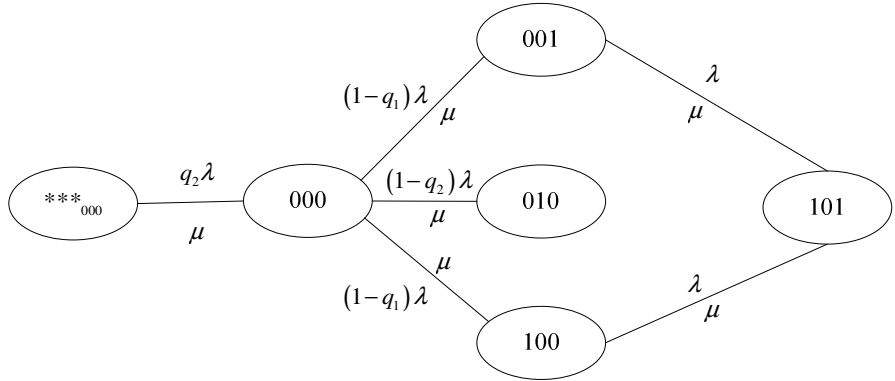

Fig. 3 Three-link network and the associated state-transition diagram with collisions

Recall that in ICN the throughput is calculated as $Th_1 = Th_3 = \dfrac{\rho+\rho^2}{1+3\rho+\rho^2}=0.7440$, $Th_2=\dfrac{\rho}{1+3\rho+\rho^2}=0.1171$ where $\rho = 83/15.5$. After taking collision effects into account,

$$Th_1 = Th_3 = \dfrac{(1-q_1)\rho+(1-q_1)\rho^2}{1+3\rho-2q_1\rho+(1-q_1)\rho^2}$$

$$Th_2 = \dfrac{(1-q_2)\rho}{1+3\rho-2q_1\rho+(1-q_1)\rho^2}$$

and collision probability of both links are

$$p_1 = p_3 = \dfrac{q_1\rho}{\rho+(1-q_1)\rho^2}, \quad p_2 = q_2$$

Hence,

$Th_1 = 0.7374$
$Th_2 = 0.1090$, $p_1 = p_3 = 0.01$
$Th_3 = 0.7374$ $\quad P_2 = 0.1174$

2) A four-link network

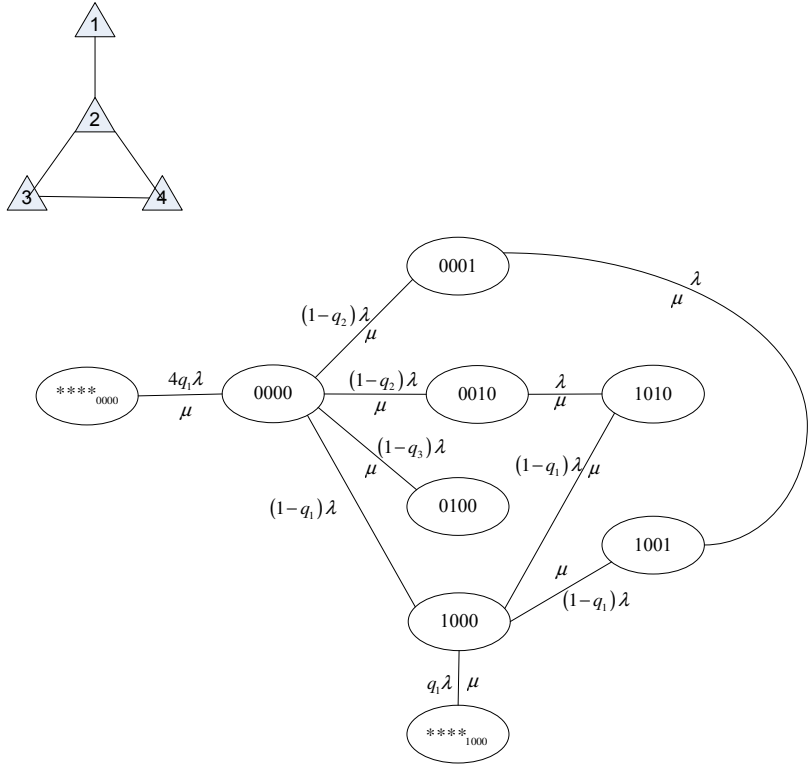

Fig. 4 Four-link network and the associated state-transition diagram with collisions

Recall that in ICN the throughput is calculated as $Th_1 = \frac{\rho + 2\rho^2}{1 + 4\rho + 2\rho^2} = 0.7861$, $Th_2 = \frac{\rho}{1 + 4\rho + 2\rho^2} = 0.0671$ $Th_3 = Th_4 = \frac{\rho + \rho^2}{1 + 4\rho + 2\rho^2} = 0.4266$ where $\rho = 83/15.5$. After taking collision effects into account,

$$Th_1 = \frac{(1-q_1)\rho + 2(1-q_2)\rho^2 + (1-q_1)q_1\rho^2}{1 + 4\rho - 4q_1\rho + 2(1-q_2)\rho^2 + (1-q_1)q_1\rho^2}$$

$$Th_2 = \frac{(1-q_3)\rho}{1 + 4\rho - 4q_1\rho + 2(1-q_2)\rho^2 + (1-q_1)q_1\rho^2}$$

$$Th_3 = Th_4 = \frac{(1-q_2)\rho + (1-q_2)\rho^2}{1 + 4\rho - 4q_1\rho + 2(1-q_2)\rho^2 + (1-q_1)q_1\rho^2}$$

$$p_1 = \frac{q_1\rho}{\rho + 2(1-q_2)\rho^2 + (1-q_1)q_1\rho^2}$$

$$p_2 = q_3$$

$$p_3 = p_4 = \frac{q_2\rho + q_1(1-q_1)\rho^2}{\rho + (1-q_2)\rho^2 + (1-q_1)q_1\rho^2}$$

and collision probability are

Hence,

$Th_1 = 0.7807$
$Th_2 = 0.0606$
$Th_3 = 0.4093$
$Th_4 = 0.4093$

$p_1 = 0.0056$
$P_2 = 0.1709$
$P_3 = P_4 = 0.07$

# V  Simulation Results

We conduct simulations to verify the accuracy of GICN as well as ICN. We use an ICN-simulator written with MATLAB to simulate the CSMA protocol with collisions caused by simultaneous transmissions. The link throughputs computed by ICN and GICN are compared with that obtained from the ICN-simulator.

As can been seen from Fig. 5, the throughputs and collision probability computed by GICN are very close to simulation results. On the other hand, the link throughputs do not change largely when collision effects are taken into account. That is, the original ICN model yields good approximations even though it does not capture the collision effect.

| Topology | ICN | GICN | | Simulation | |
|---|---|---|---|---|---|
| | Throughput (Mbps) | Throughput (Mbps) | Collision Prob | Throughput (Mbps) | Collision Prob |
| 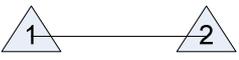 | (3.3058, 3.3058) | (3.20, 3.20) | (0.0607, 0.0607) | (3.187, 3.19) | (0.0603, 0.0604) |
| 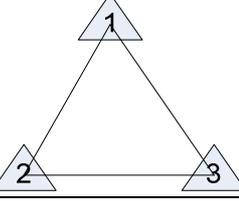 | (2.2684, 2.2684, 2.2684) | (2.12, 2.12, 2.12) | (0.1174, 0.1174, 0.1174) | (2.1208, 2.122, 2.1196) | (0.1177, 0.1174, 0.1171) |
| 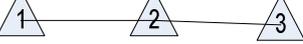 | (5.3782, 0.8463, 5.3782) | (5.3306, 0.788, 5.3306) | (0.01, 0.1174, 0.01) | (5.3263, 0.792, 5.3273) | (0.0102, 0.1178, 0.0101) |
| 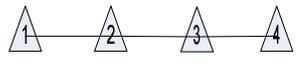 | (4.1799, 2.2684, 2.2684, 4.1799) | (4.1145, 2.1575, 2.1565, 4.1145) | (0.033, 0.07, 0.07, 0.033) | (4.1114, 2.1603, 2.1555, 4.1192) | (0.033, 0.07, 0.0691, 0.0323) |
| 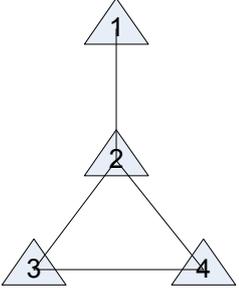 | (5.6825, 0.4853, 3.0839, 3.0839) | (5.6434, 0.4375, 2.9592, 2.9592 | (0.0056, 0.1709, 0.07, 0.07) | (5.6399, 0.4385, 2.9553, 2.961) | (0.0055, 0.1723, 0.0698, 0.0699) |
| 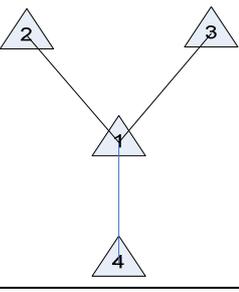 | (0.1478, 5.9669, 5.9669, 5.9669) | (0.1302, 5.9576, 5.9576, 5.9576) | (0.1709, 0.0016, 0.0016, 0.0016) | (0.1306, 5.9574, 5.9587, 5.9568) | (0.1717, 0.0017, 0.0016, 0.0016) |

Fig. 5 Contention graphs of various network topologies and the corresponding ICN, GICN computed link throughputs and simulation results.

To justify our Assumption 1 in Section III, we conduct a set of simulations, comparing the link throughputs and link collision probability with or without contention-window doubling upon collisions. As shown in Fig. 6, when we disable the contention-window doubling, the collision probability is a bit higher. However, the link throughputs remain almost the same.

| Topology | With window doubling | | Without window doubling | |
|---|---|---|---|---|
| | Throughput (Mbps) | Collision Prob | Throughput | Collision Prob |
| 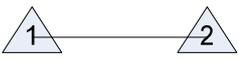 | (3.177, 3.1811) | (0.0587, 0.0587) | (3.187, 3.19) | (0.0603, 0.0604) |
| 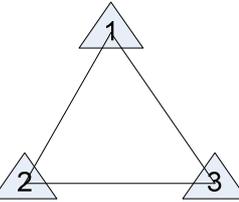 | (2.1137, 2.1165, 2.1154) | (0.1079, 0.1074, 0.1078) | (2.1208, 2.122, 2.1196) | (0.1177, 0.1174, 0.1171) |
| 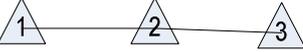 | (5.4416, 0.6644, 5.4405) | (0.0085, 0.1186, 0.0084) | (5.3263, 0.792, 5.3273) | (0.0102, 0.1178, 0.101) |
| 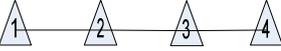 | (4.127, 2.1269, 2.1336, 4.1178) | (0.0314, 0.0673, 0.0679, 0.0318) | (4.1114, 2.1603, 2.1555, 4.1192) | (0.033, 0.07, 0.0691, 0.0323) |
| 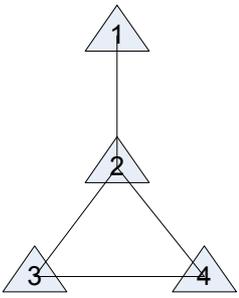 | (5.6952, 0.3835, 2.9779, 2.9714) | (0.0314, 0.0673, 0.0679, 0.0318) | (5.6399, 0.4385, 2.9553, 2.961) | (0.0055, 0.1723, 0.0698, 0.0699) |
| 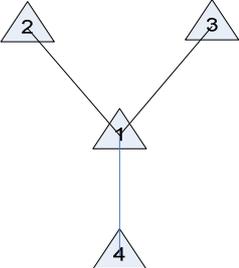 | (0.0942, 5.9926, 5.9922, 5.9918) | (0.1795, 0.0013, 0.0013, 0.0011) | (0.1306, 5.9574, 5.9587, 5.9568) | (0.1717, 0.0017, 0.0016, 0.0016) |

Fig. 6 Contention graphs of various network topologies and the corresponding simulation results with/without contention-window doubling.

# VI Conclusion

This paper presented a generalized ICN model to characterize the collision effects in CSMA wireless networks. Based on the GICN model, link throughputs and collision probability can be computed. We found that the link throughputs do not change much after taking into account the collisions incurred by simultaneous transmissions.

Several assumptions are assumed in the analysis, such as the transmission time is of the fixed length. The removal of such assumptions awaits further work.